\documentclass[prb,twocolumn,amsmath,amssymb,floatfix,superscriptaddress]{revtex4-2}
\usepackage{graphicx} 
\usepackage{booktabs}
\usepackage[dvipsnames]{xcolor}
\usepackage[version=4]{mhchem}
\usepackage{xspace, bm, braket}
\usepackage{siunitx}
\usepackage{verbatim}
\usepackage{subfigure}
\usepackage[colorlinks, citecolor=blue, linkcolor=blue, urlcolor=blue, breaklinks]{hyperref}
\usepackage{hyperref}
\include{Braket}
\usepackage{dsfont}
\usepackage[normalem]{ulem}
\usepackage{mathtools}

\newcommand{\NNO}{\ce{NdNiO2}}
\newcommand{\CCO}{\ce{CaCuO2}}
\newcommand{\dxxyy}{$d_{x^2-y^2}$}

\newcommand{\angstrom}{\textup{\AA}}

\begin{document}

\title{Superconductivity and Antiferromagnetism in NdNiO$_2$ and CaCuO$_2$: A Cluster DMFT Study} 

\author{Jonathan Karp}
\email{jk3986@columbia.edu}
\affiliation{Department of Applied Physics and Applied Math, Columbia University, New York, NY 10027, USA}

\author{Alexander Hampel}
\affiliation{Center for Computational Quantum Physics, Flatiron Institute, 162 5th Avenue, New York, NY 10010, USA}

\author{Andrew J. Millis}
\affiliation{Center for Computational Quantum Physics, Flatiron Institute, 162 5th Avenue, New York, NY 10010, USA}
\affiliation{Department of Physics, Columbia University, New York, NY 10027, USA}

\date{\today}

\begin{abstract}
We perform a comparative $2 \times 2$ real space cluster DMFT study on minimal models for NdNiO$_2$ and CaCuO$_2$ obtained from downfolding DFT states, using a Nambu formalism that allows for both superconducting and antiferromagnetic order. We produce a phase diagram in temperature and doping. We find that for the nickelate, like the cuprate, the stoichiometric compound is antiferromagnetic. We find superconductivity in a doping range bounded, with a small coexistence region, by the onset of antiferromagnetism at low doping and with transition temperature becoming immeasurably small at high doping. Superconductivity emerges at around the same hole doping for both compounds, but requires a larger deviation from half filling for the nickelate. Both antiferromagnetic and superconducting order lead to a partial gapping of the $d_{x^2-y^2}$ Fermi surface sheet. Our similar results for the cuprate and nickelate suggest that nickelate superconductivity is cupratelike. We compare our results to the experimental phase diagram. 
\end{abstract}

\maketitle

\section{Introduction}

Even over 30 years after the discovery of unconventional superconductivity in high $T_c$ cuprates, we still lack a generally accepted explanation for the mechanism of superconductivity in these materials. The recent discovery of superconductivity in the infinite layer nickelates such as \NNO{} \cite{Hwang2019, Osada2020} may shed light on this issue. This material is composed of square NiO$_2$ layers weakly coupled in the third dimension by a layer of Nd atoms \cite{Hwang2019}, and it is isostructural to the high $T_c$ cuprate \CCO{}. \NNO{} shares with cuprates the feature of the transition metal \dxxyy{} being the main active orbital with a \dxxyy{}-derived band crossing the Fermi level \cite{botana2019}. Both materials exhibit a ``superconducting dome" \cite{li2020superconducting, zeng2020phase}, a region of superconductivity in the temperature-doping plane. However, \NNO{} differs in some ways from the cuprates, most notably the Nd-derived bands that cross the Fermi level, causing a self-doping effect in the Ni-\dxxyy{} band and hybridizing with the other Ni-3$d$ bands. Additionally, the nickelate has a larger charge transfer energy than cuprates \cite{botana2019, karp2020manybody, Botana2021Nickelate}. These similarities and differences raise the question of whether the nickelate superconductivity is ``cuprate superconductivity in a nickelate'' or something more complex. 

Two experimental groups have independently measured the superconducting phase diagram of Nd$_{1-x}$Sr$_x$NiO$_2$ \cite{li2020superconducting, zeng2020phase}. They both report a superconducting state for $x$ between about $0.12$ and $0.25$ with a suppression of $T_c$ with a minimum at an $x$ between $0.15$ and $0.2$. Their results are qualitatively similar and permit comparison to our theoretical results to test the accuracy of the model and method in capturing the important contributors to superconductivity in \NNO{}. Another work measured the superconducting dome of infinite layer Pr$_{1-x}$Sr$_x$NiO$_2$ and found a similar phase diagram to \NNO{} with superconductivity between hole dopings of $0.12$ and $0.28$ \cite{Osada2020}. In the cuprate case superconductivity is generally expected to be between hole doping $x \approx 0.5$ and $0.25$ with a minimum in $T_c$ at $x = \frac{1}{8}$. 

The stoichiometric cuprates are antiferromagnetic insulators. Conversely, no long range AFM order has been found experimentally in stoichiometric \NNO{} down to $\SI{1.7}{K}$ \cite{hayward2003synthesis}. Instead, it is found to be weakly metallic \cite{li2020superconducting, zeng2020phase}. However, NMR experiments on Nd$_{0.85}$Sr$_{0.15}$NiO$_2$ find evidence of antiferromagnetic fluctuations \cite{cui2021evidence}. Some of the experimental differences between cuprate and nickelate superconductors may arise from a self doping effect due to the Nd-derived bands. Other differences may come from a difference in the underlying physics. 

Calculations based on Density Functional Theory (DFT) are in agreement that there is a main active band of mixed Ni-\dxxyy{} and O-$p_\sigma$ character crossing the Fermi level and a self doping band. The main contributors to the self doping band are Nd-$d_{z^2}$, which forms a pocket around $\Gamma$ and hybridizes with Ni-$d_{z^2}$, and Nd-$d_{xy}$, which form a pocket around the $A$ point and hybridizes with Ni-$d_{xz/yz}$ \cite{karp2020manybody}. Different groups have performed beyond DFT calculations using Dynamical Mean Field Theory (DMFT) and have come to differing conclusions. Some argue that only the Ni-\dxxyy{} orbital is important for the correlation physics \cite{karp2020manybody, karp2020comparative, Kitatani2020nickelate}, while others claims that multiple orbitals have important contributions to correlation physics \cite{wang2020hunds, kang2020infinitelayer, kang2020optical, lechermann2019late, lechermann2020multiorbital, lechermann2021doping, petocchi2020normal}. 

In terms of antiferromagnetism, single site DMFT calculations based on tight binding models fit to \NNO{} where the \dxxyy{} orbital is treated as correlated indicate that the system should be in an AFM metal phase \cite{Gu2020substantial, karp2020manybody}. Furthermore, a study based on quantum chemistry methods shows that \NNO{} has an AFM coupling of similar magnitude to \CCO{} \cite{Katukuri2020}. Another work uses a particle-hole bubble approximation to calculate the static lattice magnetic susceptibility and claims that magnetic order is frustrated \cite{leonov2020lifshitz}. 

In this paper we further investigate the hypothesis that the basic physics of cuprates and nickelates is similar by a comparative DFT+ cluster DMFT study of \NNO{} and \CCO{}. Our calculations incorporate the Nd-derived bands which both act as a charge resevoir and hybridize with the \dxxyy{} band. While there have been many DMFT studies on \NNO{} \cite{karp2020manybody, karp2020comparative, lechermann2019late, lechermann2020multiorbital, wang2020hunds, kang2020infinitelayer, kang2020optical, leonov2020lifshitz, ryee2019induced, werner2019nickelate, petocchi2020normal, si2020topotactic, Kitatani2020nickelate, Gu2020substantial, liu2020doping, lechermann2021doping}, these studies typically do not address superconductivity directly, with the exception of Ref. \cite{Kitatani2020nickelate} which uses the Dynamical Vertex Approximation (D$\Gamma$A) on top of single site DMFT to calculate the superconducting pairing susceptibility. Here, we study the superconducting state directly by performing DMFT calculations in the basis of Nambu spinors, allowing us to directly measure the anomalous Green's function in the symmetry broken phase. We perform the calculations with a real space $2 \times 2$ cluster, the minimum cluster size necessary to allow for $d$-wave superconductivity. Cluster DMFT completely takes into account temporal correlations while also considering the most important spatial correlations. We allow both superconducting (SC) and antiferromagnetic (AFM) order to study the interplay of these two types of ordering. 

\begin{figure*}[t]
    \centering
    \includegraphics[width = \textwidth]{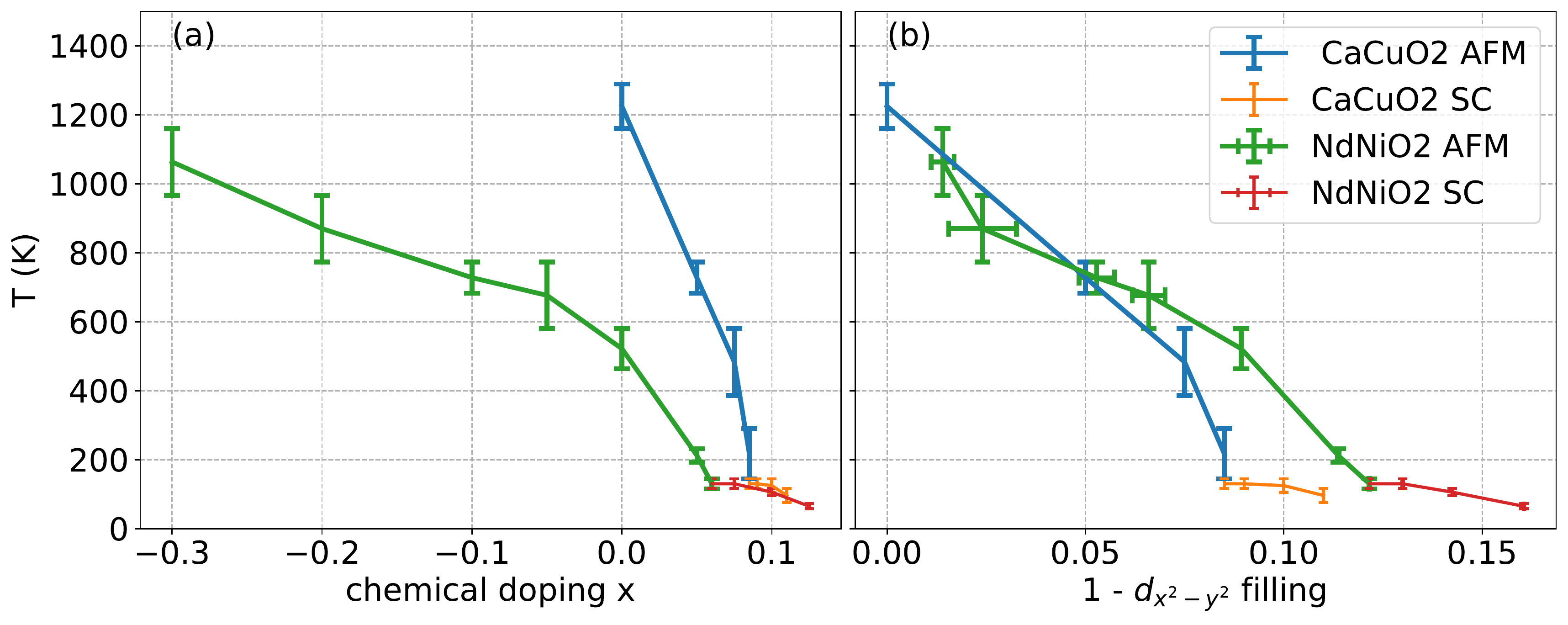}
    \caption{Phase diagram showing the $T_c$ of AFM and SC order as a function of hole doping (a) and deviation of the $d_{x^2-y^2}$ orbital from half filling (b). The points at the ends of the vertical error bars are the temperatures found to be above and below the transition temperature. Horizontal error bars represent the difference in filling at points above and below the transition. Negative $x$ corresponds to electron doping.}
    \label{fig:phase_diagram}
\end{figure*}

\section{Methods}

We perform DFT calculations using WIEN2k~\cite{Blaha2018} with the standard PBE version of the GGA functional~\cite{PBE}. For both materials we use the experimentally determined P4/mmm space group crystal structure with $a = b = \SI{3.92}{\angstrom}$ and $c = \SI{3.31}{\angstrom}$ \cite{Hwang2019} for \NNO{} and $a = b = \SI{3.86}{\angstrom}$ and $c = \SI{3.20}{\angstrom}$ for \CCO{}. The DFT calculations are converged with an $RK_{max}=7$ and with a $k$-point grid of $40 \times 40 \times 40$. The Nd-$4f$ orbitals are treated in the open core approximation. We simulate the effect of Sr doping on a DFT level  using the virtual crystal approximation (VCA), where we adjust the atomic numbers of the Nd/Ca ions to fractional values and correspondingly change the number of electrons. The VCA calculations reveal a doping dependence of the relative energies of the relevant bands, which is absent in a rigid band approximation. 

We construct our low energy models using maximally localized Wannier functions (MLWFs) \cite{MLWF1, MLWF2} using Wannier90 \cite{wannier90_v3}. For \CCO{} we use a minimal model of one Cu-\dxxyy{}-derived band, whereas for \NNO{} we construct a Wannier Hamiltonian consisting of one correlated Ni-\dxxyy{} orbital and two Nd-centered orbitals \cite{karp2020manybody}. It is necessary to keep the two Nd-derived bands for a realistic description of \NNO{} because DFT and DFT+DMFT calculations show that  bands of Nd-$d_{z^2}$ and Nd-$d_{xy}$ character cross the Fermi level and hybridize with other Ni bands \cite{botana2019, karp2020manybody}. Additionally, the electron pocket due to the Nd bands is necessary to describe the experimental change in hall coefficient sign \cite{li2020superconducting}. However, the Ni orbitals other than \dxxyy{} have small self energies \cite{karp2020comparative}, so it is reasonable to treat the Nd Wannier states as uncorrelated. For \NNO{} we use the SLWF method \cite{Wang14} to selectively localize the Ni-\dxxyy{} Wannier function. We keep all hopping matrix elements in the constructed Wannier Hamiltonians in all our calculations to properly capture the nature of the \dxxyy{} band. Further details are given in Appendix \ref{sec:wannier}.

We perform cluster DMFT calculations using the TRIQS software library \cite{TRIQS}. We use a real space $2 \times 2$ cluster in the Nambu basis. Details of how these calculations are performed are given in Appendix \ref{sec:DMFT}. We force the normal part of the self energy to have $D_4$ symmetry, but allow symmetry breaking due to AFM order. We force the anomalous self energy to have \dxxyy{} symmetry. The local Green's function is constructed using a $40 \times 40 \times 40$ $k$-point grid. 

For \NNO{} we assume only the Ni-\dxxyy{} orbitals are correlated, and for both materials we assume a site local interaction of the form:
\begin{equation}
    H_{\text{int}} = U \sum_{j = 0}^3 c^\dagger_{j \uparrow} c_{j \uparrow} c^\dagger_{j \downarrow} c_{j \downarrow}
\end{equation}
where $j$ labels the sites of the cluster, and solve the impurity problem using the continuous time hybridization expansion impurity solver CTHYB \cite{TRIQS/CTHYB}. Since we keep the Nd orbitals in the self-consistency condition for \NNO{}, we use a double counting (DC) correction, in the spirit of the fully localized limit DC, of the form $\Sigma_{dc} = U(n_{DFT} - 0.5)$, where $n_{DFT}$ is the DFT density of the Ni-\dxxyy{} orbital. We determine the static Coulomb interaction $U(\omega=0)$ of each compound using the constrained random phase approximation (cRPA) as implemented in \textsc{VASP}~\cite{Kaltak2015}, constraining the polarization function to our chosen correlated subspace. We find an onsite Coulomb interaction of the \dxxyy{} orbital of $U = \SI{2.8}{eV}$ for \NNO{} and $U = \SI{3.2}{eV}$ for \CCO{}, in agreement with Ref.~\cite{nomura2019,petocchi2020normal}. We assume that $U$ is not strongly affected by doping and use the same $U$ for each doping level.

We start each DMFT calculation with small SC and AFM seeds. For a given doping level and temperature, we classify the compound as superconducting if the anomalous self energy at the first Matsubara point $\Delta(i\omega_0)$ goes to a constant and as normal if it goes to 0. Likewise, we use the magnetization to determine whether or not the compound is antiferromagnetic. We find that reaching convergence sometimes takes at least dozens of DMFT iterations, as the number of iterations required to reach convergence increases as the phase transitions are approached.

\section{Results}

\subsection{Phase Diagram}

\begin{figure}[h]
    \centering
    \includegraphics[width = \linewidth]{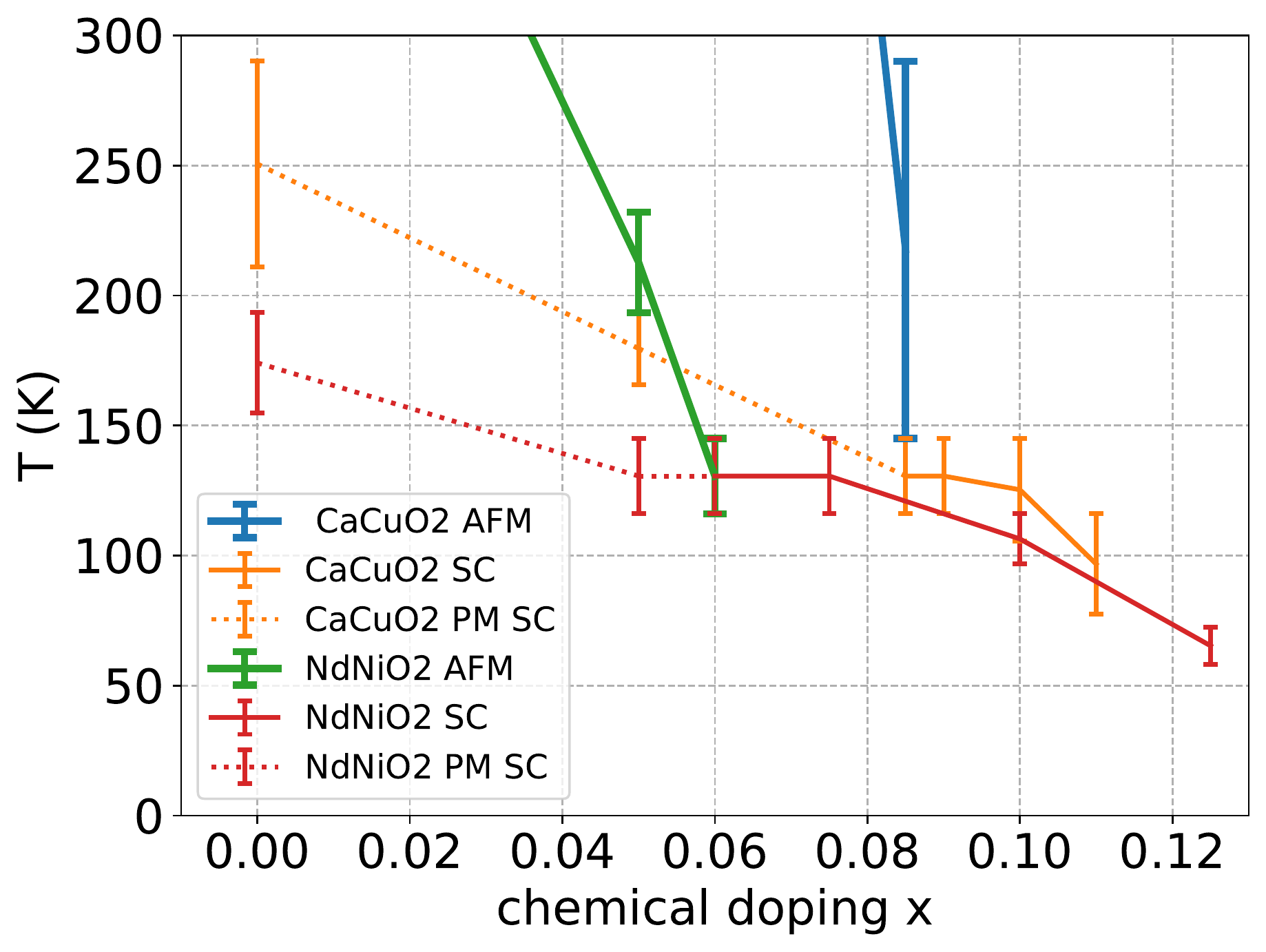}
    \caption{Phase diagram zoomed in on lower temperatures to show the superconducting $T_c$. The Superconducting $T_c$ in the case of forced paramagnetism is also shown (dashed lines).}
    \label{fig:SCTc}
\end{figure}

As described above, doping is performed using VCA in which the Nd nuclear charge in \NNO{} (or Ca nuclear charge in \CCO{}) is decreased by $x$. Figure \ref{fig:phase_diagram} panel (a) shows the AFM and SC phase transition temperatures as a function of total chemical doping $x$. Overall, the phase diagrams of \NNO{} and \CCO{} show roughly similar behavior. In the stoichiometric case, both materials exhibit a paramagnetic (PM) to commensurate (Neel) AFM transition but we do not find a SC transition. The AFM $T_c$ is much higher for \CCO{} than \NNO{}, and this remains true for small dopings. Hole doping $x$ reduces the AFM $T_c$ and for $x$ greater than a critical value the magnetism disappears. In the nonmagnetic phase, superconductivity is found. The superconducting transition is at slightly lower dopings for \NNO{} because the AFM phase is suppressed at smaller chemical dopings. Both materials exhibit a smaller coexistence region, where the transition to weak AFM and to SC appear at around the same temperature. 

Figure \ref{fig:SCTc} shows a version of Figure \ref{fig:phase_diagram} (a) zoomed in around the region of superconductivity. Also shown is the superconducting transition in the case of suppressed antiferromagnetism, where superconductivity appears for both materials even in the stoichiometric case. 

Because of the self doping of \NNO{} due to the Nd-derived bands, the \dxxyy{} orbital is already doped to a filling of about $0.91$ at stoichiometry while \CCO{} is at half filling. As $x$ is changed, the Cu-\dxxyy{} occupation is $1 - x$, and the Ni-\dxxyy{} occupation varies more slowly in a nonlinear fashion. It is therefore interesting to look at the phase diagram as a function of \dxxyy{} filling, obtained from the local impurity Green's function. Figure \ref{fig:phase_diagram} (b) shows the phase diagram as a function of deviation of the \dxxyy{} orbital from half filling. At small deviations from half filling, the AFM transition temperatures are more similar, but the \CCO{} AFM phase ends at a smaller deviation from half filling than the \NNO{} AFM phase. This may be a consequence of the larger bandwidth of \CCO{}.

The filling of the \dxxyy{} orbital obtained from the impurity Green's function may change as a function of temperature in the case of \NNO{}, especially as the onset of AFM order changes the filling. The horizontal error bars in panel (b) represent the difference between the fillings at the closest temperatures above and below the transition. The large change in filling indicates that the choice of double counting correction may be more influential in the AFM case (see appendix \ref{sec:dc}). 

Our phase diagrams show a strong competition between AFM and SC order. As soon as AFM order is suppressed by doping, a superconducting transition appears. Likewise, in the case of forced paramagnetism a superconducting transition appears even in the stoichiometric case. The competition with AFM order means that measuring the SC susceptibility could indicate the transition at a higher temperature but the actual phase transition could be lower, inhibited by AFM order. This underlines the importance of measuring the superconducting state directly while allowing for competing orders.

Comparing to experiment \cite{li2020superconducting, zeng2020phase}, we find that superconductivity starts at a doping of $\sim 0.06$ (\dxxyy{} occupancy of $\sim 0.88$) while experimentally it starts at $\sim 0.12$ (corresponding to our \dxxyy{} occupancy of $\sim 0.84$). In our calculations the low point of superconductivity is set by competition with AFM order, while experimentally no long range AFM order has been found. Additionally, the shape of the superconducting region is different from experiment, as we find a half dome instead of a full dome. A recent D$\Gamma$A study \cite{Kitatani2020nickelate} of a one band Hubbard model with a filling adjusted to fit \NNO{} finds that the onset and shape of the superconducting transition temperature depends on $U$. While their calculation incorporates antiferromagnetism, they do not explicitly construct an antiferromagnetic state and consider its competition with superconductivity. 

\subsection{Antiferromagnetism and Fermi Surface}

\begin{figure}[t]
    \centering
    \includegraphics[width = \linewidth]{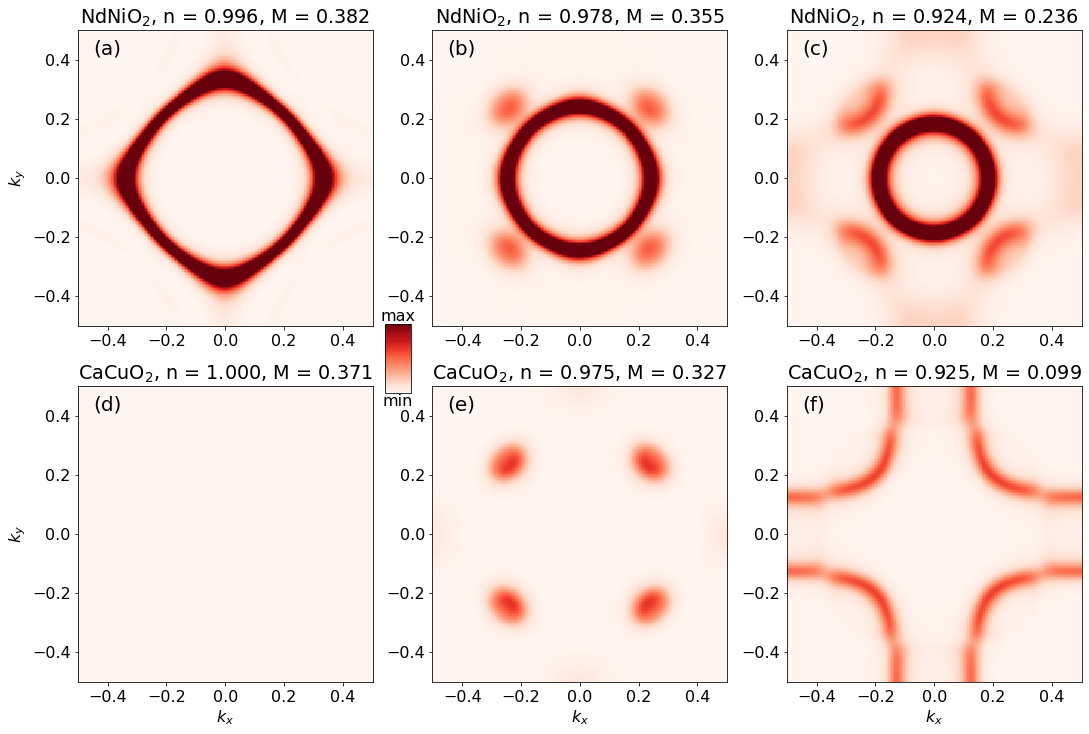}
    \caption{Many body Fermi surface in the $k_z = 0$ plane approximated by $-G(k, \tau = \beta/2)$ for \NNO{} and \CCO{} in the AFM state at different chemical dopings that give about the same $d_{x^2-y^2}$ filling $n$ for both compounds. Panel (a) corresponds to chemical doping $x = -0.2$, panel (b) corresponds to $x = -0.1$, and panel (c) corresponds to the stoichometric case of $x = 0$. For \CCO{} (d, e, and f) the doping is given by the filling. All results are for $T = \SI{290}{K}$}
    \label{fig:AFM_FS}
\end{figure}

Figure \ref{fig:AFM_FS} shows the many body Fermi surface $A(k, \omega = 0)$ in the $k_z = 0$ plane in the non-superconducting AFM case for different chemical dopings that give roughly the same average \dxxyy{} orbital occupation for both materials. Since doing analytic continuation on a matrix self energy is difficult, we approximate the Fermi surface using $-G(k, \tau = \beta/2)$ \cite{Fuchs:2011}. For the paramagnetic undoped cases, this gives a similar Fermi surface to our single site results with maximum entropy analytic continuation \cite{karp2020manybody}.

Compared to the paramagnetic case, we see that AFM has a significant effect on the Fermi surface. In the cuprate case, at half filling the Fermi surface is completely gapped. At a filling of $0.975$ the Fermi surface is gapped at the edges but not completely along the diagonal. At a filling of $0.925$ the magnetization is weaker and the Fermi surface is only slightly impacted where the extra band due to antiferromagnetism crosses the original band. In the case of \NNO{} at half filling, the Fermi surface sheet due to Ni-\dxxyy{} disappears but the Nd sheet is still present. At a filling of $0.975$ the Ni-\dxxyy{} sheet is similar to the Cu-\dxxyy{} sheet. However, the sheets are different at a filling of $0.925$ because the AFM order is stronger for the nickelate and has a greater influence in gapping the Fermi surface. 

Note that at a filling of 0.925 the \CCO{} Fermi surface is hardly affected by AFM. Yet, this doping is still below the required hole doping for superconductivity to emerge, which underscores that at this level of theory superconductivity can only emerge when AFM does not influence the Fermi surface to any large extent. 

Panel (c) is the case of stoichiometric \NNO{}. A significant part of the \dxxyy{} Fermi surface sheet is gapped, qualitatively consistent with the weak metallic behavior seen experimentally in stoichiometric NdNiO$_2$. However, the $\Gamma$-centered Fermi surface sheet due to the self doping band corresponds to weakly damped well-defined quasiparticles and should result in strong conductivity. Why this is not observed is still an open question. 

\subsection{Anomalous Self Energy}

\begin{figure}[t]
    \centering
    \includegraphics[width = \linewidth]{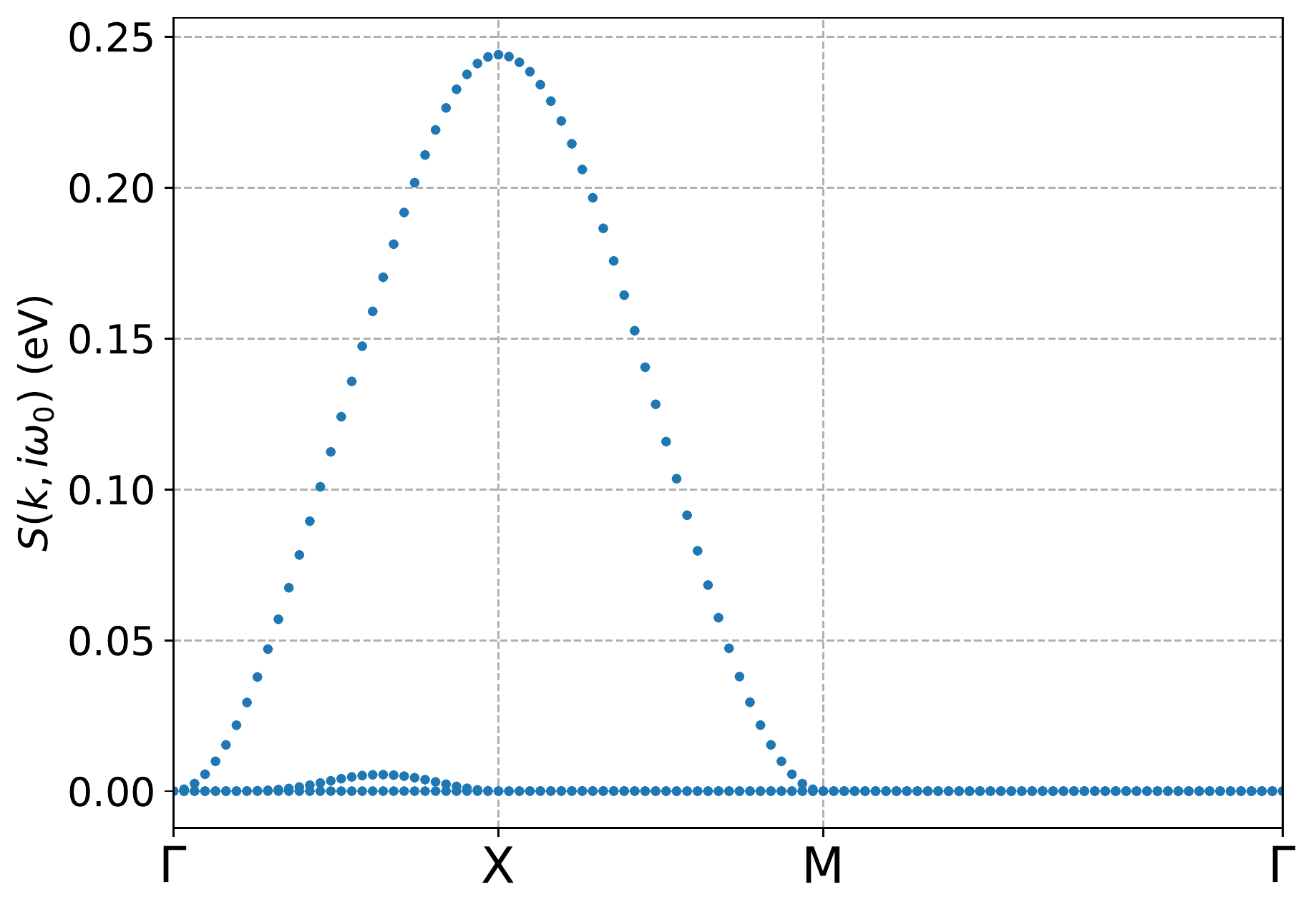}
    \caption{Band basis anomalous self energy at the first Matsubara point for undoped \NNO{} with forced paramagnetism  at $T = \SI{116}{K}$ in the full Brillouin zone.}
    \label{fig:Sigma_anomalous}
\end{figure}

The superconductivity in this model arises from interactions among electrons in the transition metal d$_{x^2-y^2}$ orbital, so the Fermi surface of the quasi two dimensional d$_{x^2-y^2}$-derived band is gapped by superconductivity.  Figure \ref{fig:Sigma_anomalous} investigates the extent to which the spectator bands inherit superconducting properties by plotting  the band basis anomalous self energy.  Despite the hybridization with Nd bands, almost all of the superconductivity is in the \dxxyy-derived band, with only a tiny part going to one of the Nd bands where it hybridizes with Ni. Since this model has \dxxyy{} symmetry, the superconducting order vanishes along the Brillouin zone diagonal where $k_x = k_y$.

Our result of very weak but $d$-symmetry superconducting order in the spectator bands is in some tension with a recent experimental report of a superconducting state with both $d$ and $s$-wave components, which are attributed to  the \dxxyy{} and the Nd-derived bands respectively \cite{Gu2020Single}. Obtaining such a result within our current theory would require including attractive interactions on the Nd orbitals. 

We further study superconductivity on the self doping band by using a Hartree approximation for the superconductivity induced on the $\Gamma$ pocket from the \dxxyy{} band self energy. We work within a model where Wannier functions for the other Ni-$d$ orbitals are added to the tight binding model, bringing the total number of bands to seven, but only the Ni-\dxxyy{} orbital is treated as correlated (for details see Appendix \ref{sec:7band}). We then calculate the anomalous self energy on the $\Gamma$ pocket band as:

\begin{equation}
    S_{\mu}(k) = -\frac{1}{N_q} \sum_q U_{\mu \nu}(k,q) F_\nu(q)
\end{equation}
where $S_\mu(k)$ is the anomalous self energy on the Nd-$d_{z^2}$ derived self doping band (labeled $\mu$), $F_\nu(q)$ is the anomalous Green's function of the \dxxyy{}-derived band (labeled $\nu$), and the interband pairing amplitude arising from the Ni-$d$ interaction is:

\begin{equation}
    U_{\mu \nu}(k,q) = \sum_{\alpha \beta \gamma \delta} w_{\alpha \mu}^*(k)w_{\beta \mu}^*(-k)w_{\gamma \nu}(q) w_{\delta \nu}(-q) U_{\alpha \beta \gamma \delta}
\end{equation}
where $w_{\alpha \mu}(k)$ is the $\alpha$ orbital element of the eigenvector corresponding to band state $\mu$ and $U_{\alpha \beta \gamma \delta}$ is the four index coulomb interaction tensor in orbital space. In this case, we take $U_{\alpha \beta \gamma \delta}$ to take the usual Kanamori form \cite{Kanamori1963} with just the Ni \dxxyy{} and $d_{z^2}$ orbitals treated as correlated with $U = \SI{2.8}{eV}$ and $J = \SI{0.7}{eV}$. 

\begin{figure}[t]
    \centering
    \includegraphics[width = \linewidth]{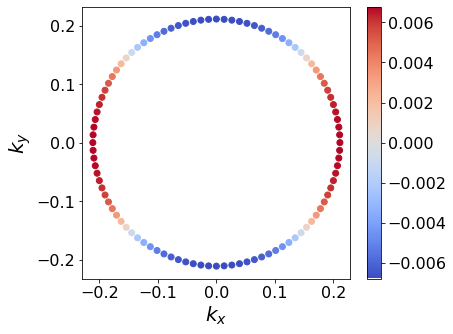}
    \caption{Anomalous self energy on the Nd $d_{z^2}$-derived $\Gamma$ pocket at $k_z = 0$ induced from a Hartree interaction with the correlated \dxxyy{} band.}
    \label{fig:gamma_pocket_sc}
\end{figure}

Figure \ref{fig:gamma_pocket_sc} shows the resulting anomalous self energy on the gamma pocket. The self energy has a \dxxyy{} symmetry and comparison with Figure \ref{fig:Sigma_anomalous} shows that the maximum is more than an order of magnitude smaller than that of the \dxxyy{} band. 

\begin{figure}[t]
    \centering
    \includegraphics[width = \linewidth]{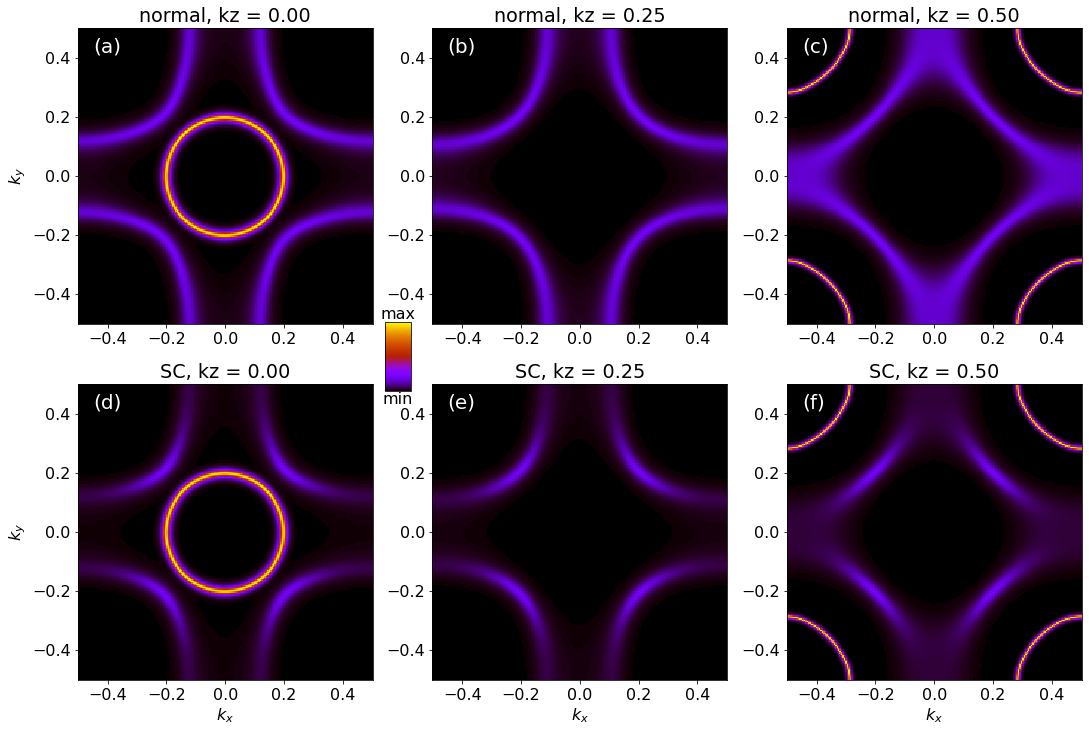}
    \caption{Comparison of the Fermi surface with only the normal self energy (top panels) and with the anomalous self energy (bottom panels) for undoped \NNO{} in the paramagnetic state at at $T = \SI{116}{K}$. The Fermi surface is approximated by $-G(k, \tau = \beta/2)$.}
    \label{fig:SC_FS}
\end{figure}

Figure \ref{fig:SC_FS} compares the Fermi surface obtained with just the normal self energy (top) and with the anomalous self energy included (bottom). Superconductivity gaps the Fermi surface along the zone edges but does not affect the zone diagonal. Comparison to Figure \ref{fig:AFM_FS} shows that both superconductivity and antiferromagnetism predominantly affect the Fermi surface near the zone edges, explaining their competition

\subsection{Increased Ni-Nd hopping}

\begin{figure}[h]
    \centering
    \includegraphics[width = \linewidth]{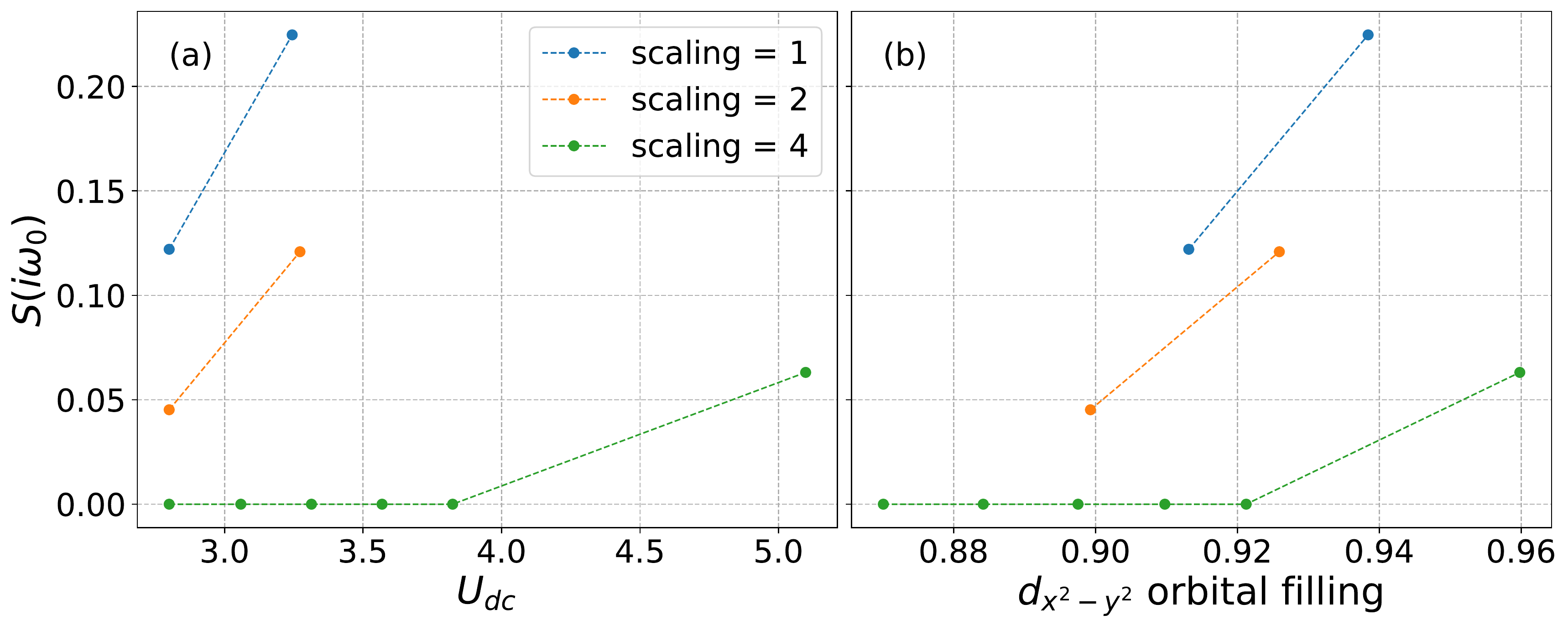}
    \caption{Representative component of the anomalous self energy at the first Matsubara point as a function of (a) $U_{dc}$ and (b) \dxxyy{} orbital filling for different values of scaling the largest Ni-Nd hopping.}
    \label{fig:SC_vs_scaling}
\end{figure}

In the three band Wannier Hamiltonian obtained from DFT calculations, the Nd-derived ``spectator" bands are relatively weakly hybridized with the Ni-d$_{x^2-y^2}$ band. The largest hopping is between the Ni orbital and the Nd-$d_{z^2}$ orbital in the neighboring cell  and is  $\SI{0.024}{eV}$, more than an order of magnitude smaller than the Ni-Ni or Nd-Nd hoppings.  Figure ~\ref{fig:SC_vs_scaling}  investigates the hypothetical case of stronger Ni-Nd hybridization by rescaling this Ni-Nd hopping by a factor of 2 and 4. 
We also vary the magnitude $U_{dc}$ of the double counting correction to ensure that different results on superconductivity are not just due to differences in \dxxyy{} orbital filling resulting from the increased Ni-Nd hopping. 

Panel (a) of Figure \ref{fig:SC_vs_scaling} shows that at fixed double counting increasing the hybridization reduces the amplitude of  the  anomalous self energy at the lowest positive Matsubara frequency, and sufficiently large interband hybridization suppresses superconductivity entirely. Panel (b) shows that increasing U$_{dc}$ also changes the $d_{x^2-y^2}$ occupancy, shifting the system out of the superconducting dome. However we further observe that even at fixed $d_{x^2-y^2}$ occupancy increasing the hybridization weakens the superconductivity. 
Increasing the hybridization also increases the anomalous self energy on the spectator bands (Figure \ref{fig:Sigma_ndx4} in Appendix \ref{sec:Ndapp}). 
These results point to the possibility of exploiting the dependence on hybridization with Nd to be able to tune superconductivity in nickelates by an external parameter that affects Ni-Nd hybridization, e.g. applying strain.

\section{Discussion}
We use cluster dynamical mean field methods to construct superconducting and antiferromagnetic states and investigate their interplay, in theoretical models downfolded from DFT calculations and believed to be relevant to the layered $d^9$ cuprate and nickelate superconducting compounds. The essential assumption of our work is that in both materials the important beyond DFT correlations arise from an onsite interaction controlling the occupancy of the transition metal d$_{x^2-y^2}$ orbital. The important chemical difference between the materials is the presence of additional spectator (``self-doping") bands, which absorb charge from the Ni d$_{x^2-y^2}$ orbitals and shift the relationship between the chemical doping and the relevant electron occupancies. 

In our work we explicitly construct the superconducting and antiferromagnetic states. This is important because we find that on the low doping side of the phase diagram, superconductivity in both materials is limited by competition with long range ordered antiferromagnetism. If long range antiferromagnetic order is suppressed, superconductivity is found in our calculation to continue to much lower dopings even though the state, within the DMFT approximation, has substantial antiferromagnetic correlations. However, an interesting closely related work based on the dynamical vertex approximation \cite{Kitatani2020nickelate}, which includes fluctuations in both channels but does not construct ordered states finds that the superconducting state terminates at a low doping independent of the onset of antiferromagnetic order when $U$ is chosen appropriately. Understanding the origin of the differences, and investigating the $U$ dependence of the different phase boundaries predicted by the different methods is an important open question.

In the NdNiO$_2$ material we also investigate the implications of adding Nd-derived spectator bands for superconductivity, showing that for parameters obtained from DFT calculations the spectator bands are essentially  decoupled, serving only as a charge reservoir, and exhibiting extremely weak superconducting properties themselves. However, increasing the hybridization with the spectator orbitals by factors of two or more strongly suppresses the superconductivity.

An important prediction of this class of theories is that the stoichiometric and lightly hole doped nickelate materials are antiferromagnetic. In the actual materials no firm evidence of antiferromagnetism has been presented, although evidence of strong antiferromagnetic exchange interactions and strong antiferromagnetic correlations has been reported \cite{zhao2021intrinsic, Lu2021magnetic}. The self-doping effect implies that chemically stiochiometric NdNiO$_2$ has many-body physics equivalent to $x\approx 0.08$-doped cuprates. It is possible that the current methods overestimate antiferromagnetism in both materials and that for example a calculation employing larger clusters would lead to a reduced antiferromagnetic range.  A clear prediction of the current methods is that {\em electron-doping} \NNO{} would lead to stronger antiferromagnetism.

Our work highlights the importance of investigations into the properties of the spectator bands including the mass enhancement and the superconducting gap amplitude. Interesting recent tunnelling measurements have been interpreted in terms of different gaps on the different bands \cite{Gu2020Single}. Further investigation, including direct determination, e.g. via photoemission, of mass enhancements and of the magnitude  and symmetry of a superconducting gap on the spectator bands would be of great interest.

The filling of the spectator bands affects the occupancy of the d$_{x^2-y^2}$ orbitals, which in turn strongly affects the physics. This filling is controlled by the double counting correction. These effects are investigated in  Appendix \ref{sec:dc} and found to be relatively minor, but further investigation is important.  

A further important open issue is the influence of other Ni-$d$ orbitals. On a DFT level, the Ni-$d_{z^2}$ orbital hybridizes with the Nd-$d_{z^2}$ and is therefore not completely filled. This had led to some authors claiming that at least two correlated orbitals are necessary to explain the relevant physics of \NNO{} \cite{lechermann2019late, lechermann2020multiorbital, petocchi2020normal, wang2020hunds, kang2020infinitelayer, kang2020optical}. A four-site cluster DMFT calculation of antiferromagnetism and superconductivity in a two-orbital model is presently not feasible, but single-site investigations of antiferromagnetism and spectator-band mass enhancements would be worth pursuing, to determine the effect of multiorbital physics on low energy properties.

Our results point to superconductivity in \NNO{} being cupratelike, regardless of differences like nickel instead of copper, the larger charge transfer energy, and the presence of self doping bands. These results refine the picture of what is important for cuprate superconductivity, which may shed more light on the mechanism. Additionally, the presence of the self doping bands may give a way of tuning the superconductivity, potentially making it easier to understand and more useful technologically.

\section{Acknowledgements}

We thank F. Lechermann and N. Wentzell for helpful discussions. The Flatiron Institute is a division of the Simons Foundation.

\appendix

\section{Wannier Functions}

\label{sec:wannier}

The MLWF procedure does not converge well for \NNO{} at some dopings, so we use the SLWF method \cite{Wang14}, selectively localizing only the Ni-\dxxyy{} Wannier function. While using SLWF instead of MLWF makes a substantial difference in the wide window \cite{Karp2021dependence}, we do not expect it to be important here because the Ni-\dxxyy{} orbital does not mix significantly with the Nd orbitals. Indeed, for hole dopings of $0.075$ and $0.1$ holes per unit cell the MLWF method converges without issue and we find that DMFT calculations with the MLWF and SLWF models give the same results. For example, at hole doping of $0.075$, the MLWF method gives a Ni-\dxxyy{} Wannier function of spread $\SI{2.743}{\angstrom}$ and the SLWF method a spread of $\SI{2.739}{\angstrom}$, and the DMFT solutions show the same behavior. We use a $k$-point grid of $21 \times 21 \times 21$ to construct the Wannier functions.

\section{Cluster DMFT Calculations}

\label{sec:DMFT}

For our 4 site cluster, the creation operators can be written as a the 4 vector:

\begin{equation}
{\bf c}^\dagger_{k\sigma} = (c^\dagger_{k \sigma 0}, c^\dagger_{k \sigma 1}, c^\dagger_{k \sigma 2}, c^\dagger_{k \sigma 3})
\end{equation}
where the sites are ordered as (0,0,0), (0,1,0), (1,0,0), (1,1,0). 
The Nambu spinors are defined as: 

\begin{equation}
{\bf \psi}_k = 
\left( \begin{matrix} {\bf c}_{k\uparrow} \\ {\bf c}^\dagger_{-k\downarrow} \end{matrix} \right) \quad
{\bf \psi}^\dagger_k = (\begin{matrix} {\bf c}^\dagger_{k\uparrow} && {\bf c}_{-k\downarrow} \end{matrix})
\end{equation}

In terms of Nambu operators, the local interaction Hamiltonian becomes:

\begin{equation}
    H_{int} = U \sum_{j = 0}^3 ( \psi_j^\dagger \psi_j - \psi_j^\dagger \psi_j \psi_{j+4}^\dagger \psi_{j+4})
\end{equation}

where $\psi_j = c_{j \uparrow}$ and $\psi_{j+4} = c_{j \downarrow}^\dagger$ are the local Nambu operators for cluster site $j$.

The Green's function in the Nambu basis is \cite{Georges1996}:

\begin{equation}
\mathcal{G}_k(\tau) = - \langle T_\tau \psi_k(\tau) \psi_k^\dagger(0) \rangle 
 = \left( \begin{matrix} G_{k\uparrow}(\tau) && F_{k\uparrow}(\tau) 
    \\ F^*_{k \uparrow}(\tau) && -G_{-k \downarrow}(-\tau) \end{matrix} \right)
\end{equation}
where $ G_{k\uparrow}(\tau) = - \langle T_\tau {\bf c}_{k \uparrow}(\tau) {\bf c}_{k \uparrow}^\dagger(0) \rangle $ is the normal Green's function

and $ F_{k\uparrow}(\tau) = - \langle T_\tau {\bf c}_{k \uparrow}(\tau) {\bf c}_{-k \downarrow}(0) \rangle $ is the anomalous Green's function. 

Fourier Transforming the Nambu Green's function gives:

\begin{equation}
    \mathcal{G}_k(i\omega_n)  = \left( \begin{matrix} G_{k\uparrow}(i \omega_n) && F_{k\uparrow}(i\omega_n) 
    \\ F^*_{k \uparrow}(-i\omega_n) && -G_{-k \downarrow}(-i\omega_n) \end{matrix} \right)
\end{equation}

Assuming that $F^*_{k \uparrow}(-i\omega_n) = F_{k\uparrow}(i\omega_n)$, we find a noninteracting Green's function of:

\begin{equation}
    \left[ \mathcal{G}^0_k(i\omega_n) \right]^{-1} = \left( \begin{matrix} i \omega_n - H(k) && 0 \\ 0 && i \omega_n + H(k)^T \end{matrix} \right) 
\end{equation}
where $H(k)$ is the noninteracting Hamiltonian in the Wannier basis.
and a self energy of the form:

\begin{equation}
    \hat{\Sigma}_{\text{Nambu}}(i \omega_n) =  \left( \begin{matrix} \Sigma_{\uparrow}(i\omega_n) && S(i\omega_n) \\ S(i\omega_n) && -\Sigma_\downarrow^*(i\omega_n) \end{matrix} \right) 
\end{equation}

where $\Sigma(i\omega_n)$ is the normal self energy and $S(i\omega_n)$ is the anomalous self energy, which is 0 in the normal state and nonzero in the superconducting state. 

Assuming $D_4$ symmetry broken by AFM order, the normal self energy has the form:

\begin{equation}
\Sigma_\uparrow(i \omega_n) = \left( \begin{matrix}
a_1 && b && b && c_1  \\
b && a_2 && c_2 && b \\
b && c_2 && a_2 && b \\
c_1 && b && b && a_1
\end{matrix} \right)
\end{equation}
with $\Sigma_\downarrow(i \omega_n)$ the same  but with $a_1$ and $a_2$ switched and $c_1$ and $c_2$ switched. Assuming superconducting order with \dxxyy{} symmetry, the anomalous self energy has the form:
\begin{equation}
S(i \omega_n) = \left( \begin{matrix}

0 && \Delta && -\Delta && 0  \\
\Delta && 0 && 0 && -\Delta \\
-\Delta && 0 && 0 && \Delta \\
0 && -\Delta && \Delta && 0
\end{matrix} \right)
\end{equation}

Using CTHYB directly in the site basis gives an average Monte Carlo sign of around 0, so we must perform a change of basis. We transform the 8x8 matrix impurity $G_0$ to a basis with states:

\begin{align*}
s_+ = 0.5 \left( \begin{matrix}1\\ 1\\ 1\\ 1\\ 0\\ 0\\ 0\\ 0\end{matrix} \right) \quad
s_- = 0.5 \left( \begin{matrix}0\\ 0\\ 0\\ 0\\ 1\\ 1\\ 1\\ 1\end{matrix} \right) \\
d_+ = 0.5\left( \begin{matrix}1\\ -1\\ -1\\ 1\\ 0\\ 0\\ 0\\ 0\end{matrix} \right) \quad
d_- = 0.5 \left( \begin{matrix}0\\ 0\\ 0\\ 0\\ 1\\ -1\\ -1\\ 1\end{matrix} \right) \\
p_{x+} = 0.5 \left( \begin{matrix}1\\ -1\\ 1\\ -1\\ 0\\ 0\\ 0\\ 0\end{matrix} \right) \quad 
p_{x-} = 0.5\left( \begin{matrix}0\\ 0\\ 0\\ 0\\ 1\\ -1\\ 1\\ -1\end{matrix} \right) \\
p_{y+} = 0.5\left( \begin{matrix}1\\ 1\\ -1\\ -1\\ 0\\ 0\\ 0\\ 0\end{matrix} \right) \quad
p_{y-} = 0.5\left( \begin{matrix}0\\ 0\\ 0\\ 0\\ 1\\ 1\\ -1\\ -1\end{matrix} \right) 
\end{align*}

In this basis, the self energy forms three blocks: a $2 \times 2$ block in the $s_+$ and $d_+$ states, a $2 \times 2$ block in the $s_-$ and $d_-$ states, and a $4 \times 4$ block in the $p$ states, given by:

\begin{equation}
    \Sigma_{sd+} = \left( \begin{matrix}
    A && B \\ B && C
    \end{matrix} \right)
\end{equation}

\begin{equation}
    \Sigma_{sd-} = \left( \begin{matrix}
    -A^* && B^* \\ B^* && -C^*
    \end{matrix} \right)
\end{equation}

\begin{equation}
    \Sigma_{p} = \left( \begin{matrix}
    D && F && -2\Delta && 0 \\
    F && D && 0 && 2 \Delta \\
    -2\Delta && 0 && -D^* && F^* \\
    0 && 2 \Delta && F^* && -D^*
    \end{matrix} \right)
\end{equation}
where 
\begin{align*}
    A &= \frac{a_1}{2} + \frac{a_2}{2} + 2b + \frac{c_1}{2} + \frac{c_2}{2} \\
    B &= \frac{a_1}{2} - \frac{a_2}{2} + \frac{c_1}{2} - \frac{c_2}{2} \\
    C &= \frac{a_1}{2} + \frac{a_2}{2} - 2b + \frac{c_1}{2} + \frac{c_2}{2} \\
    D &= \frac{a_1}{2} + \frac{a_2}{2} - \frac{c_1}{2} - \frac{c_2}{2} \\
    F &= \frac{a_1}{2} - \frac{a_2}{2} - \frac{c_1}{2} + \frac{c_2}{2} \\
\end{align*}

Note that in the paramagnetic state $a_1 = a_2$ and $c_1 = c_2$ so $B = F = 0$.

In this transformed basis, the average Monte Carlo sign becomes more manageable. However, the sign goes to 0 as temperature is reduced, limiting the temperature region where we can extract useful information.

At each DMFT iteration, we make this change of basis and force the appropriate symmetry. Before enforcing this symmetry, we ensure that terms that should be the same by symmetry converge to the same values. 
After DMFT convergence, we symmetrize the lattice self energy. We set each off diagonal term connecting nearest neighbor sites to half the DMFT value and each term connecting diagonal sites to $\frac{1}{4}$ of the DMFT value. To extract physical results such as the Fermi surface, we unfold the Green's function to the full Brillouin zone and add a form factor to cancel out the extra bands that would appear from unfolding:

\begin{equation}
    G_{\text{phys}}(k) = \sum_{a b} e^{ik\dot(R_a - R_b)} G_{ab}(k')
\end{equation}

where $G_{\text{phys}}(k)$ is the physical Green's function in the full Brillouin zone, $a$ and $b$ are indices of the four cluster sites, $R_a$ and $R_b$ are the vectors of the cluster sites, $k'$ is the $k$ point expressed in terms of the vectors of the reduced Brillouin zone of the supercell, and $G_{ab}(k')$ is the symmetrized Green's function with four cluster sites.

To test our code, we run it on a Hubbard model with just nearest-neighbor hopping, $U = 6.2 t$, a hole doping of 0.02, and forced paramagnetism. We find that the superconducting transition is between $\beta t = 17$ and $\beta t = 19$, in agreement with previous cluster DMFT studies of the Hubbard model \cite{semon2014ergodicity}. 

\begin{figure*}[t]
    \centering
    \includegraphics[width = \linewidth]{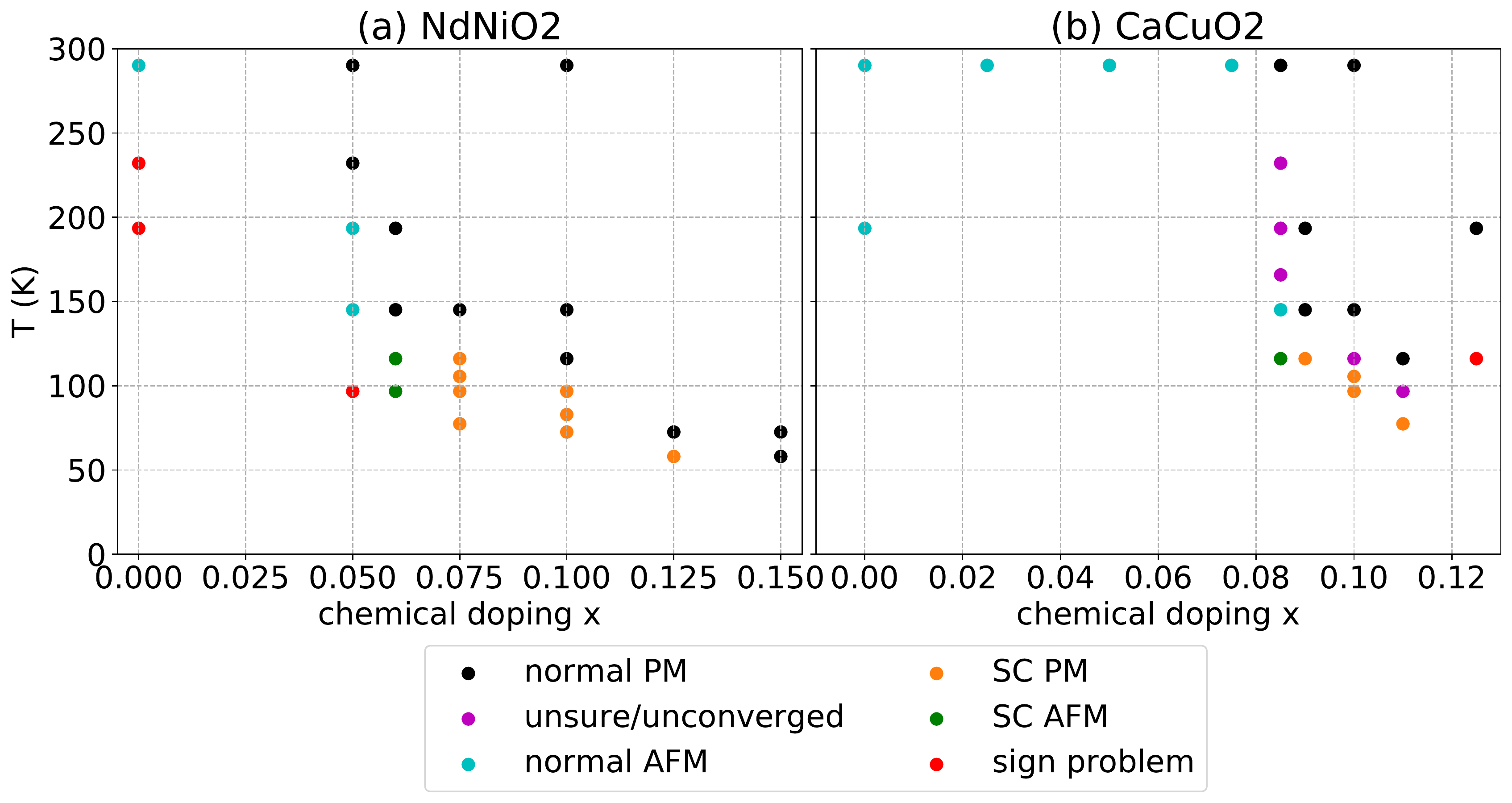}
    \caption{Phases found for all calculations done in the relevant doping and low temperature region.}
    \label{fig:calcs}
\end{figure*}

\begin{figure}[t]
    \centering
    \includegraphics[width = \linewidth]{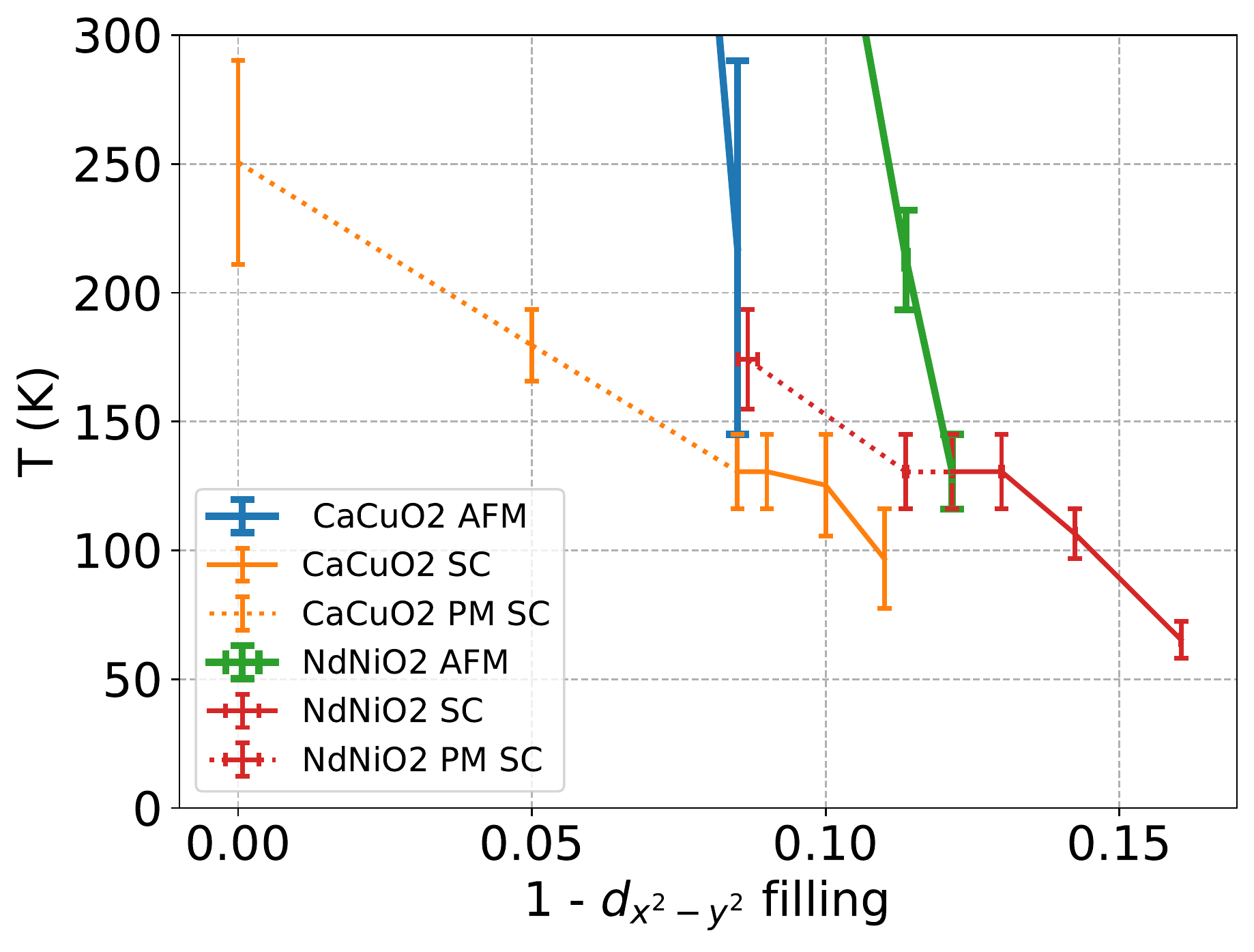}
    \caption{Phase diagram in temperature vs. deviation from half filling of \dxxyy{} orbital zoomed in on the superconducting region. The Superconducting $T_c$ in the case of forced paramagnetism is also shown (dotted lines). }
    \label{fig:SCTc_vs_filling}
\end{figure}

\section{Phase Diagram}

Figure \ref{fig:calcs} shows all of the calculations done and the resulting phases, zoomed in on low temperature and the relevant doping region for clarity. The phase boundaries of figure \ref{fig:phase_diagram} are constructed from the data here with the error bars between the highest temperature ordered point and lowest temperature unordered point. 

It is difficult to pin down the transition temperatures exactly, as the number of DMFT iterations required for convergence increases as the transition temperature is approached. Additionally, particularly in the case of superconductivity, the order parameter is small near the transition and difficult to resolve within numerical accuracy, leading to some uncertainty in the transition temperature.

\section{double counting correction}

\label{sec:dc}

One of the major uncertainties in DFT+DMFT is the double counting correction, both the functional dependence on the density and which density to use~\cite{aichhorn2011importance, Karolak:2010, hampel2020effect}. For one shot DMFT calculations, it is unclear whether it is better to use the DFT density or DMFT density. The DFT density is defined as the filling of the orbitals on a DFT level, and the DMFT density is the density obtained from the Green's function at the previous DMFT iteration. The question can be viewed as whether DFT or DMFT is ``blamed" for the double counting. In the case of fully charge self consistent DFT+DMFT, the DFT occupations of the impurities lose their physical meaning and it is apparent that the DMFT density is the correct choice. While fully charge self consistent calculations would be preferable for this reason, they would make these calculations even more computationally demanding. 

\begin{figure}[h]
    \centering
    \includegraphics[width = \linewidth]{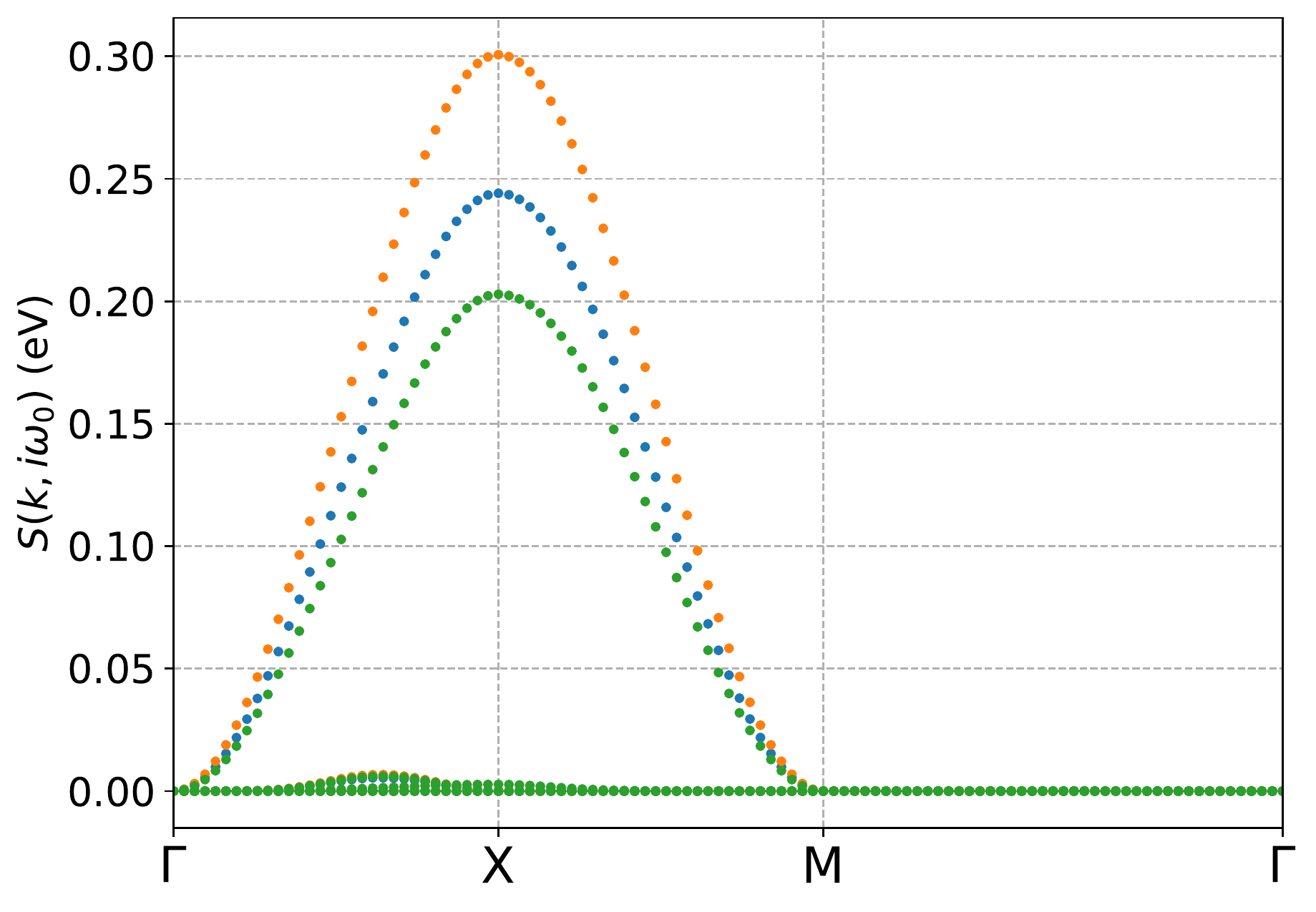}
    \caption{Comparison of the band basis anomalous self energy in three different cases. The blue is the same as figure \ref{fig:Sigma_anomalous}. The orange is the same but with the DMFT density used in the double counting correction instead of the DFT density. The green is the case where a seven band Wanniner model is used instead of three. All result are for undoped \NNO{} at $T = \SI{116}{K}$.}
    \label{fig:sigma_anom_comp}
\end{figure}

In general, full charge self consistency is only qualitatively important if DMFT leads to large charge transfers, which could particularly be an issue in a case of multiple correlated atoms~\cite{hampel2020effect}. A five orbital fully charge self consistent DFT+DMFT calculation on \NNO{} shows that the orbital occupancies do not change considerably from their DFT values \cite{karp2020comparative}. Here, in the paramagnetic state we find that using the DFT density in the double counting correction leads to a DMFT filling of the \dxxyy{} orbital not so different from the DFT filling. For example, in the undoped case the \dxxyy{} orbital has a DFT density of $0.90$. Our DMFT result at $T = \SI{116}{K}$ gives a density of $0.91$, only a slight change from the DFT density. Additionally, using the DMFT density leads to a density of $0.92$, relatively close to the result from using the DFT density. Figure \ref{fig:sigma_anom_comp} shows that using the DMFT density instead of the DFT density somewhat changes the magnitude of the anomalous self energy, but does not lead to qualitative differences. We can therefore expect that the error from the double counting correction does not greatly effect the superconducting phase boundary. 

The AFM case is somewhat more complicated. Our calculations show that the \dxxyy{} orbital is pushed significantly closer to half filling in the AFM phase. This leads to more uncertainty about the accuracy of AFM results from one shot DMFT. In the undoped case at $T = \SI{290}{K}$, using both the DFT density and the DMFT density in the double counting correction give a non-superconducting AFM phase. Using the DFT density gives a \dxxyy{} filling of 0.92 and a magnetization of 0.24, while using the DMFT density gives a filling of 0.95 and a magnetization of 0.30. We see that the choice of double counting correction makes a larger difference than in the paramagnetic state. However, that there is an AFM phase in the first place should not be influenced by the \dxxyy{} orbital filling within the AFM phase, since the starting seed only has small AFM order. 

As a check, we perform single site fully charge self consistent DFT+DMFT calculations, treating both the Ni-\dxxyy{} and Ni-$d_{z^2}$ orbitals as correlated. We do the calculations both using projectors in a wide energy window from $\SI{-10}{eV}$ to $\SI{10}{eV}$ with $U = \SI{7}{eV}$ and $J = \SI{0.7}{eV}$ and in a narrow energy window from $\SI{-3.4}{eV}$ to $\SI{2.7}{eV}$ with $U = \SI{2.8}{eV}$ and $J = \SI{0.7}{eV}$, and we find an AFM solution in both cases.

\section{increased Ni-Nd hopping}

\label{sec:Ndapp}

\begin{figure}
    \centering
    \includegraphics[width=\linewidth]{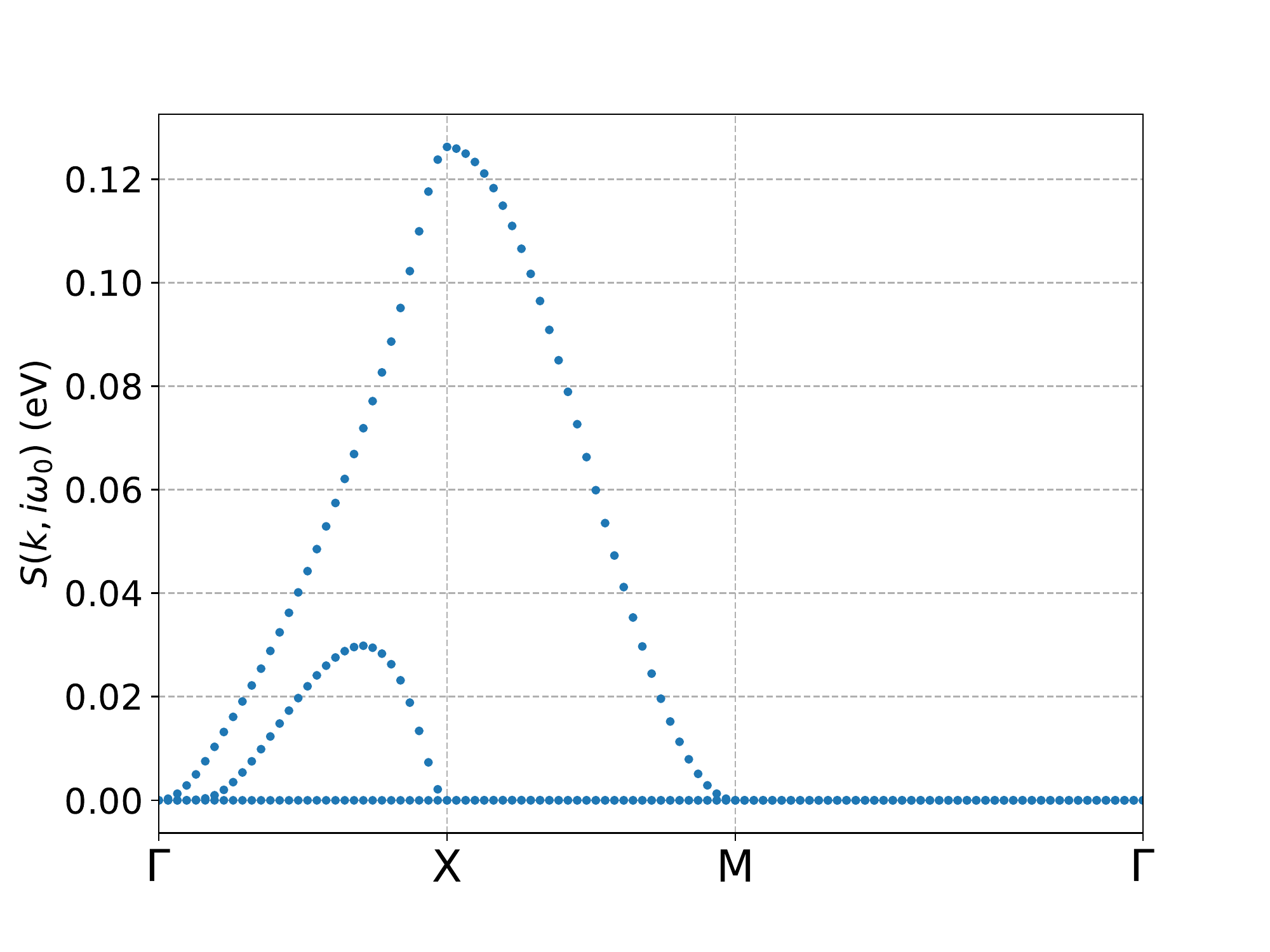}
    \caption{Same as Figure \ref{fig:Sigma_anomalous} but with the largest Ni-Nd hopping increased by a factor of 4 and the double counting correction set to $\SI{2}{eV}$ ($U_{dc} \sim \SI{5}{eV}$)}.
    \label{fig:Sigma_ndx4}
\end{figure}

Figure \ref{fig:Sigma_ndx4} shows the band basis anomalous self energy in the case with the largest Ni-Nd hopping increased by a factor of 4 and $U_{dc} \sim \SI{5}{eV}$. In that case the anomalous self energy is much more clearly on two different bands.

\section{Seven Band Model}

\label{sec:7band}

For undoped \NNO{}, we construct a seven band Wannier model, including all five Ni-$d$ orbitals and the Nd-$d_{z^2}$ and $d_{xy}$ orbitals. We perform a $2 \times 2$ cluster DMFT calculation allowing for superconductivity but not antiferromagnetism, and we keep only the Ni-\dxxyy{} orbital as correlated. We use the same double counting correction as the three band model, using the DFT density. Figure \ref{fig:sigma_anom_comp} shows the results for the band basis anomalous self energy for the 3 and 7 band models. We see that adding extra bands does not qualitatively change the anomalous self energy. This is important as it would allow a straightforward comparison to a calculation where the Ni-$d_{z^2}$ orbital is also treated as correlated. 

\clearpage

\bibliography{references.bib}

\end{document}